\documentclass[prb,twocolumn,showpacs]{revtex4-1}

\usepackage{amsfonts,amsmath,amssymb}
\usepackage{graphicx}
\usepackage{bm}

\begin{document}

%%%%%%%%%%%%%%%%%%%%%%%%%%%%%%%%%%%%%%%%%%%%%%%%%%%%%%%%%%%%%%%%%%
%%%%%%%%%%%%%%%%%%% Important commands %%%%%%%%%%%%%%%%%%%%%%%%%%%
\newcommand\rs[1]{{\scriptscriptstyle\rm #1}}
\newcommand\pdag{{\vphantom\dagger}}
\newcommand\llangle{\langle\langle}
\newcommand\rrangle{\rangle\rangle}

\title{Microwave Readout of Majorana qubits}

\author{C. Ohm}

\author{F. Hassler}
\affiliation{JARA Institute for Quantum Information, RWTH Aachen University, 52056 Aachen, Germany}

\email{ohm@physik.rwth-aachen.de}
\date{October 2014}

\begin{abstract} 
Majorana qubits offer a promising way to store and manipulate quantum
information by encoding it into the state of Majorana zero modes. As the
information is stored in a topological property of the system, local noise
cannot lead to decoherence. Manipulation of the information is achieved by
braiding the zero modes. The measurement, however, is challenging as the
information is well hidden and thus inherently hard to access. Here, we
discuss a setup for measuring the state of a Majorana qubit by employing
standard tools of microwave engineering. The basic physical effect that we
employ is the fact that a voltage-biased Josephson junction hosting Majorana
fermions allows photons to be emitted and absorbed at half the Josephson
frequency.  We show that in the dispersive regime, our setup allows us to perform
a quantum nondemolition measurement and to reach the quantum limit. An
appealing feature of our setup is that the interaction of the Majorana qubit
with the measurement device can be turned on and off at will by changing the
dc bias of the junction.
\end{abstract}


\maketitle 

%%%%%%%%%%%%%%%%%%%%%%%%%%%%%%%%%%%%%%%%%%%%%%%%%%%%%
%%%%%%%%%%%%%%%%%%%%%%%%%%%%%%%%%%%%%%%%%%%%%%%%%%%%%

\section{Introduction}

One of the major challenges of quantum computation technology is to beat
decoherence, i.e., the uncontrollable coupling of qubits to their environment.
Since the coupling of a small system to the environment is hard to control,
topological quantum computation circumvents this problem by encoding the
quantum information into global properties of a physical system and thus
making it insensitive to local perturbations.  Several ways of implementing
topological quantum computation \cite{Kitaev:2003fk,Nayak:2008zr}, ranging from
fractional quantum Hall systems at filling $\nu=5/2$ \cite{sarma:05} to
topological superconductors \cite{Kitaev:2001rt}, have been proposed.  In the
case of topological superconductors, Majorana zero modes (MZMs) (sometimes
also simply called Majorana fermions \cite{Kitaev:2001rt}) occur as mid-gap
states localized in vortex cores of two-dimensional samples or at the ends of
one-dimensional nanowires.  Most strikingly, these particles obey non-Abelian
exchange statistics, which makes them very attractive for quantum computation
applications as unitary gates can be simply applied by braiding the
MZMs \cite{Ivanov:2001ul}. A physical system in which MZMs are expected to occur
is semiconducting nanowires in proximity to a conventional superconductor in
a moderate magnetic field \cite{Lutchyn:2010qy,Oreg:2010fk,beenakker:13,Alicea:2012rt,
Leijnse:2012fr}. This system has recently attracted a lot of interest due to
experimental progress in InSb nanowires \cite{Mourik:2012qy}.

Even though the main intention of topological quantum computation is to encode
quantum information into global properties of the system, a readout for a
topological qubit can only be executed by accessing this global information
and thereby breaking the topological protection.  In the context of the
fractional quantum Hall state, it has been suggested that the readout can be
executed by interference experiments \cite{Stern:2006lr,Bonderson:2006fk}.
This was later adapted to topological superconductors where the
Aharonov-Casher effect allows to read out the Majorana qubit by interfering
fluxons \cite{Hassler:2010fj,Sau:2011kx,hassler:11,Pekker:2013lr}.

Recently, it has been reported that it is possible to couple MZMs directly to
electromagnetic radiation.  While some of the proposals primarily aimed at
identifying the signature of MZMs in microwave signals \cite{Trif:2012lr,
Cottet:2013lr, Ohm:2014ly}, others use the microwave coupling mechanism to
implement controlled qubit manipulations \cite{Schmidt:2013kx,
Schmidt:2013fj,Ginossar:yq,Yavilberg:2014}. Among different types of coupling mechanisms,
Ref.~\onlinecite{Ohm:2014ly} has shown how coherent radiation can be emitted
at half of the Josephson frequency.  The effect arises in a voltage-biased
Majorana Josephson junction and can be understood as the so-called
\emph{fractional Josephson radiation} \cite{dominguez:12}. Tuning a microwave
cavity on resonance, the Josephson junction then acts as a light source for
coherent radiation \cite{Ohm:2014ly}.
\begin{figure}[t]
  \centering
  \includegraphics[width=0.45\textwidth]{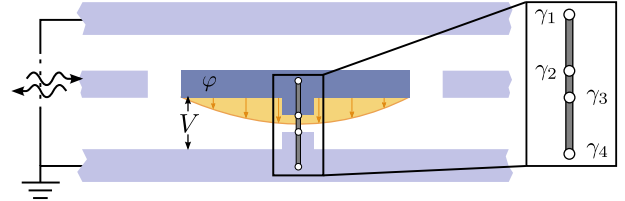} 
  \caption{The measurement setup involves three strips of superconducting
  electrodes.  The two outer strips are grounded.  The left part forms a
  transmission line (indicated by wavy-lines), whereas the darker center
  region is a strip-line microwave resonator.  A semiconducting nanowire
  bridging the center strip to the lower strip in a moderate magnetic field
  implements a Majorana Josephson junction hosting four localized MZMs
  $\gamma_1, \ldots, \gamma_4$ (white dots).  The junction is characterized by
  a time-dependent superconducting phase difference $\varphi(t)$ generated by
  a voltage $V= \hbar \dot\varphi/2e$.  The voltage $V$ consists of a dc
  component $V_0$ by which the measurement can be turned on and off and an ac
  component induced by the microwave radiation in the transmission line.}
  \label{fig:setup}
\end{figure}
In this work, we want to employ the fractional Josephson radiation in order to
implement a readout scheme for Majorana qubits.  Instead of tuning the
microwave cavity on resonance, we envision a setup in the dispersive regime
allowing for a quantum nondemolition measurement.  As the fractional
Josephson radiation arises from single-electron transport due to the presence
of MZMs, it allows for a direct readout of the Majorana qubit without
involving intermediate interference steps. Coupling of the readout device to the 
Majorana qubit can be turned on and off at will via control of the external
bias voltage.

In Sec.~\ref{sec:setup} we briefly introduce and review the coupling mechanism
that leads to fractional Josephson radiation.  We continue by introducing the
effective model for a dispersive readout in Sec.~\ref{sec:model}.  We then
study the susceptibility of the system to coherent radiation introduced by
coupling the cavity to a transmission line resonator (Sec.~\ref{sec:methods}).
Finally in Sec.~\ref{sec:readout}, we compute the measurement times for a
homodyne measurement scheme as well as for an intensity measurement showing
that both methods profit from phase coherence.  While this is always the case
for the homodyne measurement, in the present situation, the MZMs lead to
squeezing of the radiation, thereby also rendering the intensity measurement
phase-sensitive.

\section{Setup}\label{sec:setup}

A prominent way to emulate a topological superconductor in order to create
MZMs is to employ nanowires having strong spin-orbit interaction in
combination with an external magnetic field as well as proximity-induced
Cooper pairing \cite{Oreg:2010fk,Lutchyn:2010qy}. In its topological phase,
the nanowire hosts a pair of MZMs at its ends.  Being zero energy modes of
superconductors, MZMs are characterized by quasiparticle operators that are
Hermitian, $\gamma_j=\gamma_j^\dagger$, and fulfill the Clifford algebra
$\{\gamma_j,\gamma_k\}= 2 \delta_{jk}$.  In a superconductor, the fermion
number is strongly fluctuating, and only the fermion parity $\mathcal{P}=\pm 1
= (-1)^N$, with $N$ the total number of fermions, remains a good quantum
number.  In a one-dimensional topological superconductor, one can also define the
fermion parity of a topological section as the product of the two Majorana end
mode operators \cite{Kitaev:2001rt}. As the total fermion parity in a closed
system is conserved, a qubit can only be realized in a system having two
topological segments and thus four Majorana zero modes $\gamma_1, \gamma_2,
\gamma_3, \gamma_4$.  Denoting the fermion parities of the segment between
$\gamma_1$ and $\gamma_2$ ($\gamma_3$ and $\gamma_4$) by
$\mathcal{P}_{12}=i\gamma_1 \gamma_2$ ($\mathcal{P}_{34}=i\gamma_3 \gamma_4$),
the ground-state manifold is spanned by the states $|11\rangle$, $|\bar
11\rangle$, $|1\bar 1\rangle$, and $|\bar 1\bar 1\rangle$, with
$\mathcal{P}_{x} |p_{12} p_{34}\rangle = p_x |p_{12} p_{34}\rangle$, $p_x =\pm
1$, $x\in\{12,34\}$.  The states $|11\rangle$, $|\bar 1\bar 1\rangle$ are
characterized by an even number of fermions having the parity
$\mathcal{P}=\mathcal{P}_{12}\mathcal{P}_{34}=1$, whereas the states with an
odd number of fermionic particles $|1\bar 1\rangle, |1\bar 1\rangle$ have
parity $\mathcal{P}=-1$ \cite{Note1}. Most importantly, keeping the total
fermion parity fixed, the MZMs form a two-level system (Majorana qubit) with
the two states $|p_{12}, \mathcal{P} p_{12}\rangle$ distinguished by $p_{12}
=\pm1$.  The Pauli operators for the Majorana qubit are accordingly given by
$\sigma_z = \mathcal{P}_{12} = i \gamma_1 \gamma_2$ and $\sigma_x = i \gamma_2
\gamma_3$ \cite{beenakker:13}.

A single segment of a topological superconductor interrupted by a tunneling
junction (a Majorana Josephson junction), such as that formed, for example, by a
semiconducting nanowire bridging two superconductors, is a natural place to
realize four MZMs; see Fig.~\ref{fig:setup}.  Two MZMs are situated at the
ends of the nanowire, while two additional ones are formed on either side of
the tunnel junction with the overlap tunable by the phase difference across
the junction \cite{fu:08}. A distinctive feature of such a Majorana Josephson
junction is the ability to coherently transport single electrons (and not only
Cooper pairs) between the two superconductors \cite{Kitaev:2001rt,Fu:2009ys}.
Such an event changes the fermion parity on either side of the wire and thus
acts like a $\sigma_x$ on the Majorana qubit.  

In the following, we consider the situation in which such a Majorana Josephson 
junction is embedded in a strip line resonator and coupled to its resonator modes 
via a dipole interaction Hamiltonian. Hence the system can be described by
the total Hamiltonian
\begin{equation}\label{eq:total_Hamiltonian}
H = H_0 + H_{c} + H_{\rm dip}.
\end{equation}
The first term $H_0= \frac{1}{2} \hbar \delta_{\mathcal{P}} \sigma_z$ arises due 
to finite overlap amplitudes of the MZMs on each wire \cite{Ohm:2014ly}. This 
results in a small energy splitting $\hbar\delta_{\mathcal{P}}$ whose value in 
principle depends on the total fermion parity $\mathcal{P}$. 
The strip-line resonator with the resonance frequency $\omega_c$ is described 
by the Hamiltonian $H_c= \hbar \omega_c a^\dag a$, with $a$ the annihilation 
operators of the cavity mode fulfilling the canonical commutation relation 
$[a,a^\dag]=1$.  
Applying a voltage bias to the Majorana Josephson junction allows for single 
electron transfer that is accompanied by the emission/absorption of photons 
carrying the residual energy of the transition. The interaction of the Majorana 
Josephson junction with the electromagnetic environment is given by the dipole 
Hamiltonian \cite{Ohm:2014ly}
\begin{equation}\label{Eq:dipole-coupling}
	H_{\rm dip} 
  =	- \bm{d} \cdot \bm{E} = -2\hbar g \cos[\varphi(t)/ 2] (a+ a^\dagger) \sigma_x;
\end{equation}
here, $\varphi(t)$ is the superconducting phase difference across the junction,
which is related to the voltage by $V= \hbar\dot\varphi/2e$. The voltage 
consists of two parts $V= V_0 + V_\text{ac}$, with $V_0$ the dc voltage 
bias and a part due to the cavity field $V_\text{ac}= V_\text{zp} (a+a^\dag)$, 
which determines the light-matter interaction constant $g\simeq eV_\text{zp}/\hbar$.
Here the strength of the vacuum fluctuations, $V_\text{zp} = (\hbar \omega_c/C)^{1/2}$,
of the cavity is given by its total capacitance $C$ to ground. The physical 
significance of $\sigma_x$ is to implement the parity changes on either side, and 
$a^{(\dag)}$ refers to the absorption (emission) of a photon. For the sake of 
implementing a qubit readout via microwave radiation, a superconducting 
transmission line is capacitively coupled to the cavity acting as a bus for the 
information to be read out; see Fig.~\ref{fig:setup}.

\section{Model}\label{sec:model}

As explained below, the measurement setup will be active when the dc voltage
has a value $V_0 \approx \hbar\omega_c/e$.
Without the applied dc bias voltage, the superconducting phase difference 
$\varphi$ is constant up to quantum fluctuations due to the oscillator. Thus in this 
case, the first factor in Eq.~\eqref{Eq:dipole-coupling} is almost constant 
and the second term becomes ineffective as emission or absorption of a photon 
requires the energy $\hbar \omega_c$. Tuning the dc voltage close to the value 
$\hbar\omega_c/e$ leads to the superconducting phase difference of the form 
\begin{equation}\label{eq:phase_diff}
  \varphi(t) = \varphi_0 + \omega_\rs{J} t + \varphi_\text{ac},
\end{equation}
where we have introduced the Josephson frequency $\omega_\rs{J} = 2eV_0/\hbar$
and the initial phase difference $\varphi_0$. The last term originates from the cavity mode and 
is given by 
$\varphi_\text{ac}= (2eV_\text{zp}/ \hbar \omega_c) \, i(a-a^\dagger)$ since $\varphi_\text{ac}$ and 
$V_\text{ac}$ are canonically conjugate variables \cite{Note2}. 
Its magnitude can be estimated as 
$\varphi_\text{ac} \simeq  n^{1/2} e V_\text{zp}/\hbar \omega_c= (n e^2/\hbar \omega_c C)^{1/2}$, 
where $n$ denotes the number of photons in the cavity. In the following,
we assume that the capacity is large enough such that $\omega_c C \gg
n e^2/\hbar$, which corresponds to weak coupling ($n^{1/2} g\ll\omega_c$) \cite{Note3}. 
As a result, we can approximately set $\varphi(t) = \varphi_0 + \omega_\rs{J} t $.
The combined dynamics of the Majorana qubit and the cavity is described by
Hamiltonian (\ref{eq:total_Hamiltonian}), which is driven at half of the Josephson 
frequency due to the time-dependent superconducting phase in (\ref{eq:phase_diff}). 
By going to the interaction picture with respect to the Hamiltonian 
$H= \frac{1}{2}\hbar \omega_{\rs J} a^\dagger a$, the time evolution is determined 
by slowly varying variables $\tilde a=e^{i\omega_\rs{J}t/2} a$ corresponding to 
operators in a rotating frame. By neglecting off-resonant contributions, which 
appear as fast oscillating terms in the rotating frame, the Hamiltonian (\ref{eq:total_Hamiltonian}) 
maps to the time-independent quantum Rabi problem
\begin{equation}
	H = \tfrac12 \hbar\delta_{\mathcal{P}} \sigma_z + \hbar \Omega 
  \tilde a^\dagger \tilde a
		- \hbar g \sigma_x 
    ( e^{\frac{i}{2} \varphi_0} \tilde a+ e^{-\frac{i}{2} \varphi_0} 
    \tilde a^\dagger ),
\end{equation}
where $\Omega=\omega_c-\omega_{\rs{J}}/2 \ll \omega_c$ is the detuned
frequency in the rotating frame \cite{Note4}.   
In the following, we are interested in the regime of
large detuning $\Omega \gg g,\delta_\mathcal{P}$ that is characterized by weak
exchange of energy between the qubit and the cavity, therefore called the
dispersive regime \cite{Boissonneault:2009fk}. In this case, perturbation
theory in the small parameter $g/\Omega$ is a well-controlled approximation as
long as the number of photons $n$ in the cavity does not exceed the critical
value $\Omega^2/g^2$.  Using the technique of Schrieffer-Wolff transformation,
we obtain the effective Hamiltonian
\begin{align}\label{Eq:SWHamiltonian}
	H &= \hbar ( \Omega + \chi \sigma_z ) \, \tilde a^\dagger \tilde a  
  + \frac{\hbar}{2}(\delta_\mathcal{P} + \chi )\sigma_z \nonumber \\
	 &	\quad + \frac{\hbar \chi}{2} \sigma_z ( e^{i\varphi_0} \tilde a^2 +
   e^{-i\varphi_0} \tilde a^\dagger{}^2)
\end{align}
up to second order in $g/\Omega$.  Due to virtual processes, the interaction
shifts the cavity frequency and furthermore introduces the anomalous
(quadrature squeezing) terms $\propto \tilde a^2, \tilde a^\dagger{}^2$ which
arise due to counter-rotating terms of the transversal coupling \cite{Zueco:2009qf}. 
Note that the frequency shift as well as the squeezing
amplitude are determined by the parameter $\chi = -2 g^2 \delta_{\mathcal{P}} /
\Omega^2 \ll \Omega$.  Most importantly, both effects depend on the qubit
state $\sigma_z$.  To simplify the notation, we introduce the shifted cavity
frequency $\tilde{\Omega} = \Omega +\chi \sigma_z$ with a small shift compared
to the bare frequency $\Omega$.  Therefore, the qubit state can be extracted by
detecting the relative cavity shift $\pm\chi$ with respect to the undressed
cavity frequency.  For the case of homodyne detection, which we will discuss
below, the measurement will be due to the first term of
Eq.~\eqref{Eq:SWHamiltonian}. On the contrary, for the intensity measurement,
the last term will dominate.

\section{Methods}\label{sec:methods}

The measurement of the Majorana qubit is performed by observing the shift in
the cavity frequency, cf.~\eqref{Eq:SWHamiltonian}, by probing the cavity via
the transmission line.  We want to take into account that the detector is not
ideal in the sense that it does not capture every photon.  For that purpose,
we couple the resonator to two independent waveguides ($j\in\{1,2\}$), with
the idea that the photons in the waveguide with $j=1$ are measured while the
other photons are lost.  The transmission lines are modeled by the Hamiltonian
\begin{equation}
  H_{\rm{tl}} =\hbar\sum_{j} \int\!\frac{d\omega}{2\pi} \,
 \omega b^\dagger_j(\omega) b^{\vphantom{\dagger}}_j(\omega) 
\end{equation}
with operators $b_j^\dagger(\omega), b^{\vphantom{\dagger}}_j(\omega)$
creating and annihilating microwave photons at the frequency $\omega$
fulfilling the commutation relation $[b^\pdag_j(\omega), b^\dag_k (\omega')]=
2\pi \delta_{jk} \delta(\omega-\omega')$. The cavity field is coupled to the
waveguides by the Hamiltonian
\begin{equation}
  H_{\kappa} = i \hbar \sum_{j}\kappa_j^{1/2} 
\int\!\frac{d\omega}{2\pi}\, [a^\dagger b^\pdag_j(\omega)  - b^\dag_j(\omega)
a];
\end{equation}
here, the coupling parameters $\kappa_j>0$ denote the decay rate of the cavity
photons in the $j$-th transmission line.  In the following, we will introduce
the efficiency $\eta= \kappa_1/(\kappa_1 + \kappa_2)$ of the detector which
denotes the fraction of detected photons.  Neglecting spin-flip errors induced
by off-resonant interaction of the qubit with the radiation \cite{Note5}, the
combined qubit-cavity Hamiltonian $H$ commutes with $\sigma_z$ rendering the
measurement to be a quantum nondemolition measurement. In this case, the interaction 
of the resonator with the transmission lines can be described by the quantum Langevin 
equations \cite{Gardiner:2010fk}
\begin{align}\label{Eq:Langevin}
 \frac{d\tilde a}{d t} 
 &=  i [H + H_\kappa,\tilde a]/\hbar - \tfrac12 \sum_j \kappa_j \tilde a
 \\
 &= -i\tilde{\Omega} \tilde a  - 
 i \chi \sigma_z e^{-i \varphi_0} \tilde
 a^\dagger 
 -  \sum_{j} \Bigl[
 \kappa_j^{1/2} \, \tilde 
     a_{{\rm in},j}  + \tfrac12   \kappa_j \tilde a \Bigr],\nonumber
\end{align}
where the input field, defined as
\begin{equation}\label{Eq:input-operator}
  \tilde a_{{\rm in},j}(t) = \int \!\frac{d\nu}{2\pi}e^{- i \nu t} 
b_{j}(\omega_{\rs{J}}/2 + \nu) ,
\end{equation}
satisfies the commutation relation $[\tilde a^\pdag_{\text{in},j}(t), \tilde
a^\dag_{\text{in},k} (t')] = \delta_{jk} \delta(t-t')$.  The output field is
then given by the boundary condition $\tilde a_{\rm out,1}(t) = \tilde a_{\rm
in,1}(t) + \kappa_1^{1/2} \tilde a(t)$ at the interface to the transmission
line 1.  The stochastic differential equations~(\ref{Eq:Langevin}) can be
solved for $\tilde a$ by going to frequency space with $\tilde a(\nu)= \int
\!dt\,e^{i \nu t} \tilde a(t)$.  The solution is given by the linear relation
\begin{multline}\label{Eq:output-operator}
\tilde a_{\rm out,1}(\nu ) = \sum_{j} 
\Bigl[u_j(\nu) \tilde a_{{\rm in},j}(\nu)  + v_j(\nu) 	
  \tilde a^\dag_{{\rm in},j} ( - \nu) \Bigr]
\end{multline}
between the input and the output fields. In our case, the annihilation
operator of the output field is a coherent superposition of  annihilation  as
well as creation operators of the input field. This can be traced back to the
anomalous terms in (\ref{Eq:SWHamiltonian}). All the relevant information
about the scattering of microwaves is encoded into the functions
\begin{subequations}
\begin{align} 
u_1(\nu) &= \frac{(\frac{1}{2}\kappa_2 - i\nu)^2 - (\frac{1}{2}\kappa_1 - i\tilde{\Omega})^2-\chi^2}
			{(\frac{1}{2}\kappa-i\nu)^2+\tilde{\Omega}^2-\chi^2}, \\ 
v_1(\nu) &= \frac{i\kappa_1 \chi e^{-i\varphi_0}}
			{(\frac{1}{2}\kappa-i\nu)^2+\tilde{\Omega}^2-\chi^2}, \\
u_2(\nu) &= - \frac{\sqrt{\kappa_1 \kappa_2} \left[\frac{1}{2}\kappa- i(\nu +\tilde{\Omega})\right]} 
			{(\frac{1}{2}\kappa-i\nu)^2+\tilde{\Omega}^2-\chi^2}, \\ 
v_2(\nu) &= \frac{i  \sqrt{\kappa_1 \kappa_2}  \chi e^{-i\varphi_0}}
			{(\frac{1}{2}\kappa-i\nu)^2+\tilde{\Omega}^2-\chi^2},
\end{align}
\end{subequations}
where $\kappa=\kappa_1+\kappa_2$ is the total line width of the cavity.  Note
that they satisfy the identity $\sum_j ( |u_j|^2-|v_j|^2 )=1$ which translates
the canonical commutation relation from $\tilde a_\text{in}$ to $\tilde a_\text{out}$.

In the following, we will only consider the situation where the resonator is fed
by a coherent state at frequency $\nu_\text{in}$ with $\langle
\tilde a_\text{in,1}\rangle = e^{-i\nu_\text{in} t} \alpha_\text{in}$.  For the
second transmission line, we take the vacuum state as an input state that is
valid in the low temperature limit at temperatures $T \ll \hbar\omega_c/k_B$. 
The radiation reflected back into the first transmission line, which will be
subsequently measured, is characterized by its mean output signal
\begin{align}\label{Eq:Expectation-value}
	\left\langle \tilde a_{\rm out,1}(\nu) 
  \right\rangle &=  u_1(\nu) \alpha_{\rm in} \delta(\nu-\nu_{\rm in})  
	 + v_1(\nu) \alpha_{\rm in}^* \delta(\nu + \nu_{\rm in}) 
 \end{align}
as well as by the correlation functions
\begin{align}
	\llangle \tilde a_{\rm out,1}^\dagger(\nu) a^{\vphantom{\dagger}}_{\rm out,1}
  (\nu') \rrangle
&=	2\pi\,
|v_1(\nu)|^2 \, \delta(\nu -\nu'), \label{Eq:Correlation-functions} \\
	\llangle  \tilde a_{\rm out,1} (\nu) \tilde a_{\rm out,1} (\nu') \rrangle
&=	2\pi \,u^*_1(\nu) v^*_1(-\nu) \, \delta(\nu+\nu'); 
\nonumber
\end{align}
here and below, the double brackets denote the (co-)variance defined as
$\llangle AB \rrangle = \langle A B\rangle - \langle A \rangle \langle B
\rangle$. The squeezing term of Eq.~\eqref{Eq:SWHamiltonian} leads to the fact
that both correlators in \eqref{Eq:Correlation-functions} are nonzero.

\section{Majorana qubit readout}\label{sec:readout}

The qubit readout proceeds via the measurement of the ac voltage component
$\propto a_{\rm out,1}$ that is reflected back from the cavity into the
transmission line 1.  As the ac voltage component oscillates at microwave
frequencies, the signal frequency needs to be downconverted, which can be
achieved by means of a homodyne measurement technique as well as by measuring
the intensity which is the squared modulus of the voltage.  Both schemes will
be discussed in the following.  To keep the discussion simple, we will discuss
only the case in which the frequency of the input signal is given by $\nu_{\rm
in}= \Omega + \chi$, i.e, it is on resonance with the cavity given that the
qubit is in the state $\sigma_z = 1$.

\subsection{Homodyne detection}
\begin{figure}[t]
  \centering
  \includegraphics[width=0.25\textwidth]{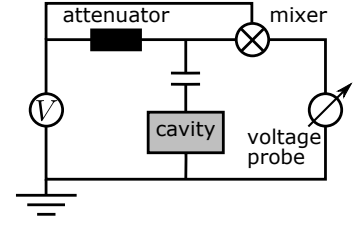} 
  \caption{Electric circuit for implementing the homodyne readout: The gray
  box symbolizes the microwave cavity of Fig.~\ref{fig:setup}.  The cavity is
  coupled via the capacitance $C$ to the microwave signal generated by the
  voltage source and subsequently attenuated (black box).  The output signal
  of the resonator is then mixed with the input signal to down-convert the
  frequency and in the end the dc component is measured.  It is important that
  while the input signal is attenuated the local oscillator strength remains
  unsuppressed therefore dominantly contributing to the output of the mixer.}
  \label{fig:homodyne}
\end{figure}
The standard homodyne measurement technique converts high-frequency signals
down to a zero frequency signal by mixing the voltage to be measured, $V_{\rm
out} = V_\text{zp} (a^\pdag_\text{out,1} + a^\dag_\text{out,1})$, with a local
oscillator, $V_{\rm lo} = V_\text{zp} \mathop{\rm Re} ( \alpha_\text{lo}e^{-i
\omega_\rs{J} t/2-i \nu_\text{in}t} )$, where both voltages oscillate at the
same frequency $\omega_\rs{J}/2 + \nu_\text{in}$; see Fig.~\ref{fig:homodyne}.
The mixer outputs the intensity $I_{\rm hd} = (V_{\rm lo} + V_{\rm out})^2$,
which can be subsequently measured at the voltmeter.  We assume that due to
the attenuator, the amplitude of the local oscillator is much larger than the
output of the cavity, $|\alpha_{\rm lo}| \gg |\alpha_{\rm in}|$.  In this
case, the leading contribution to $I_\text{hd}$ is the mixed term $2V_{\rm
lo}V_{\rm out}$ as the product $V_\text{lo}^2$ carries no information about
the qubit state.  For the measurement of the qubit state $\sigma_z = \pm 1$,
it is thus necessary to distinguish the intensities $I_\text{hd}|_{\sigma_z
=\pm 1}$ corresponding to the fact that the qubit is in one of the two
states.  Introducing the difference of the intensities $\delta I_\text{hd}=
I_\text{hd}|_{\sigma_z=1} - I_\text{hd}|_{\sigma_z=\bar 1}$ as our signal
strength, we derive the result
\begin{align}\label{eq:with_phase}
|\delta I_{\rm hd}| 	
&\simeq  \eta |\alpha_{\rm in}| |\alpha_{\rm lo}| \, \left| \frac{\chi}{\Omega} \right|
		\Bigl| 16 \frac{\Omega}{\kappa} \sin(\varphi_{\rm lo} - \varphi_{\rm in}) \nonumber \\
& \quad + (\omega_{\rm J}/\omega_c-1)^{1/2} 
		\Bigl[ \frac{\kappa}{\Omega} \sin(\varphi_{\rm lo} - \varphi_{\rm in} + \varphi_0) \nonumber \\
    & \qquad - 4 \cos(\varphi_{\rm lo} - \varphi_{\rm in} + \varphi_0) \Bigr]
    \Bigr| \\
& \lesssim \eta |\alpha_{\rm in}| |\alpha_{\rm lo}| \, \left|
\frac{\chi}{\kappa} \right| \label{eq:without_phase}
\end{align}
valid to first order in $\chi/\Omega \ll 1$. Note that the intensities, as
is typical for homodyne detection, are dependent on the phase
$\varphi_\text{in/lo}$ of the corresponding signals with $\alpha_\text{in/lo}=
|\alpha_\text{in/lo}| e^{i \varphi_\text{in/lo}}$. In going from
\eqref{eq:with_phase} to \eqref{eq:without_phase}, we have optimized the phase
$\varphi_\text{lo}$ so as to have the maximum signal. To remain in
the dispersive limit taking the cavity broadening $\kappa$ into account, 
we have to require that $\kappa/\Omega \ll 1$ such that the first
term of \eqref{eq:with_phase} that originates from the first term in 
Eq.~\eqref{Eq:SWHamiltonian} is dominating and thus gives in the optimal 
case $\varphi_{\rm lo} - \varphi_{\rm in}=\tfrac{\pi}{2}$.

In order to determine the measurement time $T$, we have to compare the
magnitude of the  signal $S_{\rm hd}= \delta I_{\rm hd} \, T$ to the noise
\begin{equation}\label{eq:noise}
N^2_{\rm hd}
=\iint_0^T\!dt\,dt'\,  \llangle I_{\rm hd}(t')I_{\rm hd}(t)\rrangle 
\simeq  T |\alpha_{\rm lo}|^2;
\end{equation}
here, we have used the fact that $|\alpha_\text{lo}| \gg |\alpha_\text{in}|$
such that the noise is dominated by the local oscillator. The minimal
measurement time $T_0$ is given by a signal-to-noise ratio $S_{\rm hd}/
N_{\rm hd} =1$. Employing Eqs.~\eqref{eq:without_phase} and \eqref{eq:noise},
we obtain
\begin{align}
  T_{0,\rm hd} &\simeq 
  \frac{\kappa^2}{\eta^2  |\alpha_{\rm in}|^2 \chi^2}.
			   \label{Eq:Homodyne-Measurement-Time}
\end{align}
Note that the measurement time is inversely proportional to the number of
photons $|\alpha_\text{in}|^2$ which are sent in per unit of time.  This is a
common behavior for measurement setups at low temperatures that are limited
by shot noise. The factor $\chi/\kappa$ which enters quadratically is there
due to the fact that we have to distinguish the qubit induced frequency shift $\chi$ 
relative to the spectral broadening $\kappa$ of the cavity. In order
to determine the minimal measurement time, we have to remember that the number
of photons in the cavity has to be smaller than the critical value
$\Omega^2/g^2$ in order for the Schrieffer-Wolff approximation to be
applicable. This translates to the bound $|\alpha_{\rm in}|^2 \lesssim
\Omega^2 \kappa / \eta g^2 $ on the input field strength and thus to
\begin{equation}\label{eq:t_hd_min}
  T_\text{0,hd} \gtrsim  \frac{\kappa g^2}{\eta \Omega^2  \chi^2}
  \simeq \frac{\kappa}{\eta \delta_\mathcal{P} \chi}.
\end{equation}
As we did not assume any relation between $\kappa$ and $\chi$, the measurement
time can be made arbitrarily small by decreasing $\kappa/\chi$.

\subsection{Intensity measurement}

Another route to achieve down-conversion is to mix the signal with itself,
i.e., to measure the intensity $I_\text{int}=V_{\rm out}^2$.  Calculating the
difference of intensities between the two qubit states yields
\begin{align}
|\delta I_{\rm int}| 	
&\simeq  \eta |\alpha_{\rm in}|^2 \, \frac{|\kappa_1-\kappa_2|}{\kappa} \left|\frac{\chi}{\Omega}\right| \nonumber \\
&\quad 	\times \Bigl| \frac{\kappa}{\Omega} \sin(\varphi_0 + 2\varphi_{\rm in}) 
				- 4 \cos(\varphi_0 + 2\varphi_{\rm in}) \Bigr|   \label{Eq:Intensity-Difference} \\
&\lesssim \eta |\alpha_{\rm in}|^2 \, \frac{|\kappa_1-\kappa_2|}{\kappa}
\left|\frac{\chi}{\Omega}\right| \label{eq:int_diff2}
\end{align}
to first nonvanishing order in $\chi/ \Omega$.  In going from
\eqref{Eq:Intensity-Difference} to \eqref{eq:int_diff2}, we have again chosen
the optimal value of the phases of the incoming signal which is $\varphi_{\rm
in} = \frac{\pi}{4}- \tfrac12 \varphi_0$ in the relevant limit $\kappa \ll
\Omega$.

Surprisingly, the first term in $\delta I_{\rm int}$ appears in first order in
$\chi/\Omega$ and not only to second order, as one would generically expect.
The reason for this can be traced back to the fact that the term is due to the
squeezing term in \eqref{Eq:SWHamiltonian}, which makes the intensity
measurement phase-sensitive by acting as a parametric amplifier operated below
the threshold; cf.\ Ref.~\onlinecite{Ohm:2014ly}. In this way, in our setup
the intensity measurement itself is phase sensitive and not only the homodyne
detection.  To zeroth order in $\chi/\Omega$, the noise is given by
\begin{equation}
	N_{\rm int}^2
  \simeq T | \, \alpha_{\rm in}|^2 |u_1(\nu_\text{in})|^2
  \simeq T | \, \alpha_{\rm in}|^2  \frac{(\kappa_1 - \kappa_2)^2}{
  \kappa^2}.
\end{equation}
Comparing the noise to the signal leads to the expression
\begin{align}
	T_{0,\rm int}	&\simeq   \frac{\Omega^2}{\eta^2 |\alpha_\text{in}|^2\chi^2}
  \gtrsim \frac{g^2}{\eta \kappa \chi^2}
  \simeq \frac{\Omega^2 }{\kappa^{2}}  T_\text{0,hd}
					\label{Eq:Intensity-Measurement-Time}			
\end{align}
for the minimal time of measurement.  Even though the intensity measurement in
our case is better than in a typical situation without parametric driving,
comparing it to the homodyne detection, we conclude that due to the additional
(large) factor $\Omega^2/\kappa^2$ the homodyne detection scheme is always
more efficient.

\subsection{Decoherence}

As long as the external bias voltage is switched off, there is no coupling to
the electromagnetic field, but nevertheless the Majorana qubit as an open
quantum system may suffer from uncontrollable interaction with its
environment.  Due to a finite overlap of MZMs on the nanowire, the qubit may be
affected by dephasing errors $\propto \delta_{\mathcal{P}}\sigma_z$ as well as
by external tunneling of quasiparticles onto the
junction \cite{Rainis:2012lr}. The latter process is also called quasiparticle
poisoning (QP).  An erroneous interaction $\propto \sigma_z$ results in
dephasing of the qubit with an intrinsic dephasing rate $\Gamma_{\delta}
\simeq \delta_{\mathcal{P}}^{-1}$, whereas the dephasing caused by QP arises due
to global parity switches \cite{budich}.
QP of the MZM can also be generated by driving-induced transitions to the 
quasiparticle continuum above the gap \cite{Houzet:2013}. We assume this mechanism 
to be negligible here as the transparency of the Josephson junction is considered 
to be relatively small. Other sources of decoherence such as
nonequilibrium fluctuations of the nanowire's chemical potential \cite{Konschelle:2013uq} or
thermal effects \cite{Cheng:2012fk,Schmidt:2012lr}, may also be included, but
here we want to focus only on QP and effects due to finite splitting. 

We wish to describe QP by a simple model where each of the mid gap states $|p_{12}
p_{34}\rangle$ can be changed to an arbitrary state in the opposite parity
sector via quasiparticle tunneling processes.  For simplicity, it is assumed
that all these processes happen with the same rate $\Gamma_{\rs{QP}}$.  The
density matrix fulfills the Lindblad equation
\begin{equation}\label{Eq:Lindblad}
 \frac{d\rho}{dt} = -\frac{i}{\hbar} [H,\rho] 
 					+ \sum_{i=0}^8 
          \Gamma_{i} \left( J^{\vphantom{\dagger}}_i \rho J_i^\dagger 
						- \frac{1}{2} \{ J_i^\dagger J^{\vphantom{\dagger}}_i, \rho \} \right)
\end{equation}
with jump operators $J_i$ implementing dephasing due to finite energy
splitting of MZMs ($i=0$) and QP-induced jump processes $i\in \{1,\dots,8\}$.
In particular, we have for the intrinsic dephasing $\Gamma_0= \Gamma_\delta$, $J_0=
\sigma_z$. The QP is modeled by $\Gamma_i =
\Gamma_\rs{QP}, i\geq 1$ with the jump operators $J_i$ given by the eight
possibilities to change the parity, e.g., $J_1= |1 1 \rangle \langle
\bar 1 1|, \dots$ , see Ref.~\onlinecite{Ohm:2014ly}.

In equation (\ref{Eq:Lindblad}), the diagonal elements decouple, which results
in an exponential decay of the qubits expectation value
\begin{equation}
  \sigma_z(t)=\mathrm{Tr}[ \sigma_z \rho(t)]= e^{-2\Gamma_{\rs{QP}}t} \sigma_z(0),
\end{equation}
with $\sigma_z(0)$ denoting the expectation value of $\sigma_z$ at time $t=0$.
The dephasing time can be inferred from the correlation function $\langle
\sigma_+(\tau) \sigma_-(0) \rangle$, which, neglecting fast oscillating terms
as well as sub-leading terms of the order $\chi/\Omega$, evaluates to
\begin{align}\label{Eq:dephasing-correlation}
	\langle \sigma_+(t) \sigma_-(0) \rangle 
	&= \left\langle \exp\left[-2i \int_0^t\!dt'\, 
    \varepsilon(t')  - \Gamma_{\rs{QP}}t -\Gamma_\delta t
		\right] \right\rangle \nonumber \\
	&=	\exp\left[-2i \langle\varepsilon\rangle t  - \Gamma_\phi t
  \right].
\end{align}
The dephasing rate
\begin{equation}\label{Eq:dephasing-time}
	\Gamma_\phi 
  = \Gamma +  \frac{8\eta \chi^2 |\alpha_\text{in}|^2 }{\pi \kappa^2} 
\end{equation}
consists of two parts.  The first part given by $\Gamma=\Gamma_{\rs{QP}}
+\Gamma_{\delta}$ is dephasing due to the Lindblad equation.  The second
source of decoherence is the fluctuation of the
 instantaneous qubit frequency $\varepsilon(t) \approx \frac{1}{2}
\delta_{\mathcal{P}} + \chi \bigl[ \langle \tilde a^\dag(t)
  \tilde a(t) \rangle +\tfrac12\bigr]$ around its mean frequency 
  $\langle \varepsilon \rangle =\tfrac12 \delta_\mathcal{P} 
  + \chi (\eta |\alpha_\text{in}|^2/\kappa +\tfrac12)$ due to the fact 
that the number of photons in the cavity $\langle \tilde a^\dag(t) \tilde a(t) \rangle$ 
changes in time .
A necessary requirement for a good readout is that the dephasing time is
dominated by the measurement setup which means that
\begin{equation}\label{eq:good_meas}
  \Gamma \ll \frac{\eta \chi^2 |\alpha_\text{in}|^2}{\kappa^2} \simeq \eta^{-1} 
  T_\text{0,hd}^{-1} \lesssim  \frac{\Omega^2 \chi^2}{\kappa g^2}.
\end{equation}
In order to reach the quantum limit for the homodyne detection, the $\lesssim$
signs in \eqref{eq:good_meas} have to be equalities and $\eta=1$ such that
$\Gamma_\phi = T_\text{0,hd}^{-1}$ and thus all dephasing is due to the
measurement. Note that in the case of intensity measurement, the quantum limit
cannot be achieved as $T_\text{0,int}^{-1} \gg T_\text{0,hd}^{-1}$.

To see the limit of the homodyne detection, it is interesting to discuss the
ideal situation without QP, $\Gamma_{\rs{QP}}=0$, such that
only intrinsic dephasing mechanisms are present. In this case, the necessary
condition Eq.~\eqref{eq:good_meas} for quantum-limited measurement implies
that $\Gamma_\delta \ll \Omega^2 \chi^2/\kappa g^2$ which can be simplified to
$\chi/\kappa \gg 1$.  Thus, in order to be able to reach the quantum limit, we
need the immediately evident result that the shift of the cavity due to the
qubit state has to be much larger than the cavity broadening.  In this
parameter range, every photon carries a sufficient amount of information, which allows for a
quantum-limited measurement.

\section{Conclusions}

We have demonstrated how the embedding of a semiconducting nanowire bridging
two superconductors in a microwave cavity can be utilized for the measurement
of the Majorana qubit due to the four MZMs at the ends and the interface.  The
measurement proceeds by probing the transmission lines capacitively coupled to
the cavity.  In the dispersive regime of large qubit cavity detuning, the
system implements a quantum nondemolition measurement.  We have shown that
the dispersive frequency shift can be either detected by a homodyne or
alternatively by an intensity measurement.  Although the intensity measurement
contains an unusually large amount of coherence, it turns out not to be
sufficient in order to compete with the dephasing time of the qubit.  In
contrast, with the homodyne measurement technique one is even able to push the
measurement time towards the quantum limit.  Thus, the microwave homodyne
readout of the Majorana qubit promises to be a new scheme having the main
advantages of providing a mechanism directly coupling microwave radiation to
the topological qubit.  Furthermore, the readout process can be turned on and
off at will by simply changing the dc bias voltage.

We acknowledge financial support from the Alexander von Humboldt foundation.

\end{document}